# Minimum Variance Multi-Frequency Distortionless Restriction for Digital Wideband Beamformer

Yipeng Liu[*†], Jia Xu[†], Qun Wan[*], Yingning Peng[†]

[*] Electronic Engineering Department, University of Electronic Science and Technology of China, Chengdu, 611731, China
[†] Electronic Engineering Department, Tsinghua University, Beijing, 100084, China
E-mail: dr.yipengliu@gmail.com; xujia@mail.tsinghua.edu.cn; wanqun@uestc.edu.cn; ynpeng@mail.tsinghua.edu.cn

*Abstract*—This paper proposes a digital amplitude-phase weighting array based a minimum variance multi-frequency distortionless restriction (MVMFDR) to aviod the frequency band signal distortion in digital beamformer and too short time delay line (TDL) requirement in analoge wideband TDL array.

***Keywords-*** amplitude-phase weighting, wideband beamformer, ultra-wideband (UWB).

## I. Introduction

Wideband beamformer has a wide application in radar, communication, medical imaging, etc [1]-[5]. It involves the manipulation of signals induced on the elements of an array system or transmitted from the elements and received at a distant point from the array. Different from traditional narrowband array, wideband signal has a wide spectrum range, which makes most of the traditional narrowband array processing methods ineffective.

In a conventional narrowband beamformer the signals corresponding to each sensor element are multiplied by a complex weight to form the array output. As the signal bandwidth increases, the performance of the narrowband beamformer starts to deteriorate because the phase provided for each element and the desired angle is not a function of frequency and, hence, will change for the different frequency components [1]. Hence, Analog tapped delay line (TDL) structure was widely used in wideband array. But in wideband array, the time delay is very short. For the ultra-wideband (UWB) one, the time delay is in order of nanosecond and even sub-nanosecond. The requirement of extremely short TDL brings the difficulties in practice [5][6].

The traditional minimum variance distortionless restriction (MVDR) weight vector incorporated a single frequency is not suitable for the other frequencies of the signal. Wideband signal has a large amount of frequencies. Usually the center frequency is chosen in the MVDR model. It is always just the center frequency component of the signal that can have the desired beam pattern. But the other frequency component of the signal would have distortion in SOI's direction. Consequently the resulted array output would have large distortion of the SOI.

To deal with the above disadvantages of previous approaches, this paper adopts the traditional amplitude-phase weight array for its convenience in practice. But different from previous MVDR beamformer which is a way to enhance the signal from the desired direction by suppressing all the sources from other directions as well as the background noise [7], we impose a group of multi-frequency distortionless restrictions in the model to achieve distortionless of wideband signal. It avoids the implement difficulties resulted from wideband analog TDL, while keeps the signal of interest (SOI) distortionless in its direction of arrival (DOA). Simulation shows that the proposed digital approach has its performance advantage over traditional MVDR.

## II. Traditional MVDR Beamformer

In traditional narrowband array, assuming the signal received by a uniform linear array (ULA) with $M$ sensors is $\mathbf{x}(k)$, where $k$ is the time index, the output of beamformer is

$$y(k) = \mathbf{w}^H \mathbf{x}(k) , \qquad (1)$$

where $\mathbf{w}$ is the complex weighting vector. Expand it to get

$$y(k) = s(k)\mathbf{w}^H \mathbf{a}(\theta_0) + \sum_{j=1}^{J} i_j(k)\mathbf{w}^H \mathbf{a}(\theta_j) + \mathbf{w}^H \mathbf{n}(k) , \qquad (2)$$

where $\mathbf{a}(\theta_0)$ and $\mathbf{a}(\theta_j)$ are the signal and interferences steering vectors respectively, $s(k)$ and $i_j(k)$ are the random amplitudes of receive signal and interferences, $J$ is the number of interferences, and $\mathbf{n}(k)$ is Gaussian noise.

The formulation of the MVDR beamformer is to minimize the array output energy, subject to a linear constraint on the desired DOA, i. e.

$$J(\mathbf{w}) = \frac{1}{2}\mathbf{w}^H \mathbf{R}_{xx} \mathbf{w} , \qquad (3a)$$

$$\text{s.t. } \mathbf{w}^H \mathbf{a}(\theta_0) = 1 , \qquad (3b)$$

where $R_{xx}$ is the covariance matrix of the received signal. A Lagrange method can solve (3) and gets a close-form solution.

## III. MINIMUM VARIANCE MULTI-FREQUENCY DISTORTIONLESS RESPONSE BEAMFORMER

Here we propose a minimum variance multi-frequency distortionless response (MVMFDR) beamformer to minimize the array output energy, subject to a group of multi-frequency distortionless restrictions on all the wideband SOI's frequencies. It can be formulated as:

$$\min_{\mathbf{w}} \mathbf{w}^H \mathbf{R}_x \mathbf{w}, \tag{4a}$$

$$s.t. \begin{array}{l} \mathbf{w}^H \mathbf{a}(\theta_0, f_1) = b \\ \mathbf{w}^H \mathbf{a}(\theta_0, f_2) = b \\ \vdots \\ \mathbf{w}^H \mathbf{a}(\theta_0, f_M) = b \end{array}, \tag{4b}$$

where $f_1, f_2, \ldots, f_M$ is the discrete frequency points of the signal band, $M$ is the number of frequency points, and $b$ is a constant. (4) can be solved similarly as (3).

To do so, we rewrite (4) as

$$\min_{\mathbf{w}} \mathbf{w}^H \mathbf{R}_x \mathbf{w}, \tag{5a}$$

$$s.t. \quad \mathbf{w}^H \mathbf{A} = \mathbf{B}, \tag{5b}$$

where

$$\mathbf{A} = \left( \mathbf{a}(\theta_0, f_1) \cdots \mathbf{a}(\theta_0, f_M) \right), \tag{6}$$

$$\mathbf{B} = \left( b \cdots b \right), \tag{7}$$

and define the function $f(\mathbf{w}, \lambda_1, \ldots, \lambda_M)$ as

$$f(\mathbf{w}, \lambda_1, \cdots, \lambda_M) = \mathbf{w}^H \mathbf{R}_x \mathbf{w} - \lambda_1 \left( \mathbf{w}^H \mathbf{a}(\theta_0, f_1) - b \right) - \cdots - \lambda_M \left( \mathbf{w}^H \mathbf{a}(\theta_0, f_M) - b \right), \tag{8}$$

where $\lambda_1, \ldots, \lambda_M$ are the Lagrange multipliers and we define $\boldsymbol{\lambda} = [\lambda_1, \ldots, \lambda_M]^T$. The following set of equations then applies:

$$\frac{\partial f}{\partial \mathbf{w}} = 2\mathbf{R}_x \mathbf{w} - 2\lambda_1 \mathbf{a}(\theta_0, f_1) - \cdots - 2\lambda_M \mathbf{a}(\theta_0, f_M) = 0. \tag{9}$$

Then, we can get the solution

$$\mathbf{w} = \mathbf{R}_x^{-1} \mathbf{A} \boldsymbol{\lambda}. \tag{10}$$

Substituting (10) into (5b), we can get:

$$\boldsymbol{\lambda} = \frac{\mathbf{B}^H}{\mathbf{A}^H \mathbf{R}_x^{-1} \mathbf{A}}. \tag{11}$$

By imposing all signal frequency band distortionless restrictions in the SOI direction, the array outputs the distortionless wideband signal. And the minimization of the output energy reduces the interference as it does in MVDR.

## IV. SIMULATION

The simulation shows the performance of the proposed MVMFDR. A uniform linear array (ULA) with eight half-wavelength of the highest signal frequency spaced sensors is used. Spatially white Gaussian noise is assumed with unity variance. The actual source DOA is supposed to be 50°, and the interference DOA is set to 80°. The SOI-to-noise ratio (SNR) is set to 20 dB, and the SOI-to-interference ratio (SIR) is 1/2. 64 snapshots were used. 500 independent trials were averaged. The wideband signal frequency is in the range [3.50GHz, 3.60GHz]. The MVDR beamformer adopts the center frequency 3.55GHz to calculate the array weight, while the MVMFDR uses five distortionless restrictions on frequencies 3.50GHz, 3.52GHz, 3.55GHz, 3.57GHz and 3.60GHz. If necessary, we can impose a larger number of frequency distortionless response restrictions to tighten it. The solutions of Model (3) and (4) are calculated by convex programming software cvx [8].

Fig. 1 is the normalized beam patterns of two approaches in different frequencies. Obviously, In the SOI direction, the MVDR has different array gains in different frequency components which would result to the signal distortion. Besides, low array gain in SOI direction would lead to low received SNR. It is showed that the MVMFDR nearly overcomes the above two drawbacks. The beam patterns of different frequencies have the almost the same array gain in the SOI direction, and its level in SOI direction is higher than MVDR's while the levels of two approaches in interference direction are almost the same. Hence, MVMFDR is the most suitable one in the digital wideband beamformers.

## V. CONCLUSION

This paper proposed the MVMFDR beamformer based on traditional amplitude-phase weight array for wideband beamforming. It eliminates the distortion in traditional MVDR beamformer in processing wideband signal, and avoids the extreme short TDL requirement in wideband array. Overcoming these two disadvantages qualifies it as a better wideband beamforming, especially for UWB beamforming.

In the future, the relationship between the number of multi-frequency distortionless restrictions and beamformer's performance is going to be investigated. The theoretic performance of the proposed MVMFDR will be evaluated and the numerical experiments would be enriched too.

ACKNOLEGEMENT

This work was supported in part by the National Natural Science Foundation of China under grant 60772146, the National High Technology Research and Development Program of China (863 Program) under grant 2008AA12Z306, the Key Project of Chinese Ministry of Education under grant 109139, China National Science Foundation under Grant 60971087, China Ministry Research Foundation under Grant 9140A07011810JW0111 and 9140C130510D246, Aerospace Innovation Foundation under Grant CASC200904.

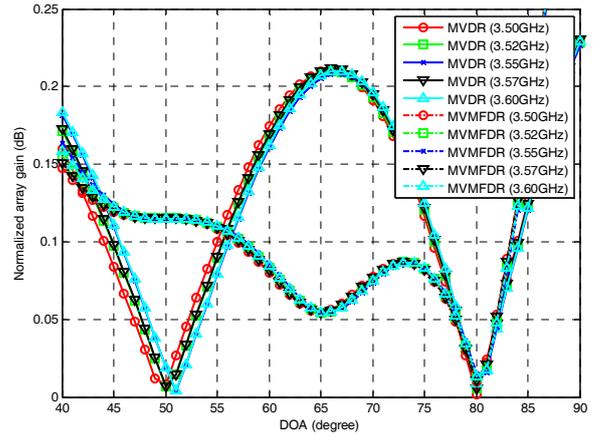

Fig. 1. Beam patterns of MVDR and MVMFDR in different frequencies.